\documentclass[aps,prd,twocolumn,nofootinbib,nobibnotes]{revtex4-1}
\pdfoutput=1

\usepackage[usenames,dvipsnames]{xcolor}
\usepackage{hyperref}
\hypersetup{
	colorlinks=true,
	urlcolor=blue,
	linkcolor=blue,
	citecolor=blue,
	bookmarks=true,
	pdftitle={A New Framework for Higher Loop Witten Diagrams},
	pdfauthor={Aidan Herderschee},
	pdfdisplaydoctitle=true,
	pdfstartview=FitH
}

\usepackage{amsmath}
\usepackage{amsfonts}
\usepackage{mathtools}
\usepackage{microtype}
\usepackage{tikz}
\usepackage{bm}
\usepackage[utf8]{inputenc}
\usepackage{blkarray}
\usepackage{bigstrut}
\usepackage{nccmath}
\usepackage{enumitem}
\usepackage{braket}
\usepackage{dsfont}
\usepackage[english]{babel}
\usepackage{empheq}

\allowdisplaybreaks

\def\ads{\textrm{AdS}}
\usepackage{tikz}
\usepackage[compat=1.1.0]{tikz-feynman}

\def \f[#1] {{\color{color#1}\bm{f_#1}}}
\def \a[#1] {{\color{color#1}\bm{\alpha_#1}}}
\def \col[#1]#2 {{\color{color#1}\bm{#2}}}
\def \cov[#1]   {\genfrac{}{}{0pt}{0}{}{\medmath{#1}}}
\def \contr[#1] {\genfrac{}{}{0pt}{0}{\medmath{#1}}{}}

\begin{document}

\begin{flushright}
LCTP-21-36
\end{flushright}


\title{A New Framework for Higher Loop Witten Diagrams}
\author{Aidan Herderschee}
\email{aidanh@umich.edu}
\affiliation{
 Leinweber Center for Theoretical Physics, \\
Randall Laboratory of Physics, Department of Physics, \\
University of Michigan, Ann Arbor, MI 48109, USA
}

\begin{abstract}\noindent
The differential representation is a novel formalism for studying boundary correlators in $(d+1)$-dimensional anti-de Sitter space. In this letter, we generalize the differential representation beyond tree level using the notion of operator-valued integrals.  We use the differential representation to compute three-point bubble and triangle Witten diagrams with external states of conformal dimension $\Delta=d$. We compare the former to a position space computation. 
\end{abstract}

\maketitle

\section{Introduction}

Boundary correlation functions in anti-de Sitter (AdS) space provide an important laboratory for studying quantum field theory and quantum gravity. The AdS background regulates possible infrared (IR) divergences in perturbation theory \cite{CALLAN1990366,Aharony:2012jf,Aharony:2015hix}, and the AdS/CFT correspondence provides a computationally tractable example of holography \cite{tHooft:1993dmi,Susskind:1994vu,Maldacena:1997re,Gubser:1998bc,Witten:1998qj,Penedones:2010ue}. This work focuses on AdS boundary correlators, for which many different computational methods have been developed \cite{Aharony:2016dwx,Alday:2017xua,Yuan:2017vgp,Giombi:2017hpr,Alday:2017vkk,Aprile:2017qoy,Alday:2018pdi,Alday:2018kkw,Yuan:2018qva,Bertan:2018khc,Liu:2018jhs, Bertan:2018afl,Chester:2019pvm,Aprile:2019rep, Carmi:2018qzm,Ponomarev:2019ofr,Carmi:2019ocp,Aprile:2019rep,Alday:2019nin,Ghosh:2019lsx,Meltzer:2019pyl,Drummond:2019hel,Meltzer:2019nbs,Albayrak:2020bso,Meltzer:2020qbr,Alday:2020tgi,Costantino:2020vdu,Carmi:2021dsn,Alday:2021ajh}. A particularly fruitful approach to searching for new methods has been generalizing established techniques for computing scattering amplitudes in flat space. For example, the OPE inversion formula \cite{Caron-Huot:2017vep,Simmons-Duffin:2017nub} is the AdS generalization of the Froissart-Gribov formula \cite{Gribov:1961ex}. 

Motivated by this approach, a new representation of AdS boundary correlators, the differential representation, has emerged. It is in a sense analogous to the momentum representation of scattering amplitudes. Momentum vectors are replaced by non-commuting conformal generators acting on a contact diagram \cite{diffflatspace}. The differential representation of AdS boundary correlators was first proposed in Refs. \cite{Roehrig:2020kck,Eberhardt:2020ewh} using the infinite tension limit of certain string theory expressions and further developed in Refs. \cite{Diwakar:2021juk,Gomez:2021qfd,Sivaramakrishnan:2021srm}.


In this letter, we generalize the differential representation of scalar AdS correlators beyond tree level by introducing the notion of operator-valued integration. We find that operator-valued integrals of scalar Witten diagrams can be interpreted as integrals over a non-commutative space. For example, operator-valued integrals obey a generalization of integration-by-parts (IBP) \cite{Chetyrkin:1981qh,vonManteuffel:2012np,vonManteuffel:2014ixa,Maierhofer:2017gsa,Mastrolia:2018uzb,Smirnov:2019qkx,Frellesvig:2019kgj,Frellesvig:2019uqt,Mizera:2019vvs,Mizera:2019ose,Frellesvig:2020qot,Laporta:2000dsw}, which is discussed in Section \ref{genintegrationidentites}. After evaluating these operator-valued integrals, the higher loop correlators in AdS become functions of conformal generators acting on contact diagrams. To illustrate the new methodoloy, we compute three-point bubble and triangle Witten diagrams in $d=2$ and $d=2,3,4$ dimensions respectively using the differential representation. We compare the former to a more traditional computation performed in position space.  To the author's knowledge, closed form expressions for the triangle Witten diagram in general dimension were previously unknown \cite{6jsymbolex}.

\section{The Differential Representation}\label{diffrep}

We begin with a brief review of the differential representation. We work in embedding space, $\mathbb{R}^{d+1,1}$, where $P^{A}$ and $X^{A}$ denote boundary and bulk coordinates respectively \cite{10.2307/1968455,deAlfaro:1976vlx,Weinberg:2010fx,Costa:2011mg}. The boundary-to-bulk and bulk-to-bulk propagators of scalars are denoted as $E_{\Delta}(P,X)$ and $G_{\Delta}(X,X')$ respectively where the $\Delta$ subscripts are suppressed if $\Delta=d$. Unless stated otherwise, the conformal dimension of all states, external and internal, is restricted to $\Delta=d$ for simplicity. For integrals over AdS boundary or bulk coordinates, we suppress the $d$ or $d+1$ superscript in the differential. Conventions are reviewed in Appendix \ref{concventionsapp}. 

The differential representation of the $n$-point correlator takes the form 
\begin{equation}
\mathcal{A}_{n}=\hat{\mathcal{A}}_{n}\mathcal{C}_{n}   
\end{equation}
where $\hat{\mathcal{A}}_{n}$ is the differential correlator, a collection of differential operators that act on a scalar contact diagram, 
\begin{equation}
\mathcal{C}_{n}=\int_{\ads}dX \prod_{i} E(P_{i},X)   \ .
\end{equation}
Scalar differential correlators, $\hat{\mathcal{A}}_{n}$, can be written solely using conformal generators, which are
\begin{equation}\label{conformalgenerat}
D_{i}^{AB}=\frac{1}{\sqrt{2}}\left ( P^{A}_{i}\frac{\partial}{\partial P_{B,i}} -P^{B}_{i}\frac{\partial}{\partial P_{A,i}}\right ) 
\end{equation}
in the embedding space formalism. Isometry generators of bulk coordinates, denoted as $D_{X}$, are the same as Eq. (\ref{conformalgenerat}) except with the replacement of the boundary coordinate $P$ with bulk coordinate $X$. While momentum space scattering amplitudes are functions on a commutative kinematic space parameterized by $p_{i}^{\mu}$, differential correlators are operator-valued functions on a non-commutative kinematic space parameterized by $D_{i}^{AB}$. The AdS analog of momentum conservation is the Conformal Ward Identity (CWI). 

%
%

An explicit example is instructive. Consider the integrand of the four-point s-channel Witten diagram
\begin{equation}\label{finalintegrandexp}
\begin{split}
\mathcal{A}_{\textrm{s-channel}}=&\int_{\ads} dX_{1}dX_{2}E(P_{3},X_{1})E(P_{4},X_{1})\\
&\quad \times G(X_{1},X_{2})E(P_{1},X_{2})E(P_{2},X_{2}) \ . 
\end{split}
\end{equation}
To derive the differential representation, we first use that
\begin{equation}\label{equationofmotion}
G(X_{1},X_{2})=-(\square_{X_{1}})^{-1}\delta^{d+1}(X_{1},X_{2}) \ ,    
\end{equation}
where $\square_{X}=-D_{X}^{2}$ is the AdS Laplacian, to rewrite the position space Witten diagram as
\begin{equation}
\begin{split}
&\mathcal{A}_{\textrm{s-channel}}=\int_{\ads}dX_{1}dX_{2} E(P_{3},X_{2})E(P_{4},X_{2}) \\
&\times E(P_{1},X_{1})E(P_{2},X_{1})\frac{1}{D_{X_{1}}^{2}}\delta^{d+1}(X_{1},X_{2}) \ . 
\end{split}
\end{equation}
We can then use the identity,
\begin{equation}
(D_{X}^{AB}+D_{1}^{AB}+D_{2}^{AB})E(P_{1},X)E(P_{2},X)=0   \ ,
\end{equation}
to replace $D_{X}^{2}$ with 
\begin{equation}
D_{12}^{2}=(D_{1,AB}+D_{2,AB})(D_{1}^{AB}+D_{2}^{AB}) \ .   
\end{equation}
Ultimately, one finds the differential representation of the s-channel Witten diagram is
\begin{equation}\label{schannelexp}
\mathcal{A}_{\textrm{s-channel}}=\frac{1}{D_{12}^{2}}\mathcal{C}_{4}\ .  \end{equation}
The differential representation of higher point Witten diagrams is analogous to Feynman diagrams under the replacement of propagators with the inverse differentials, $1/D_{I}^{2}$. For example, the five-point Witten diagram is
\begin{equation}\label{5point}
\begin{split}
\pgfmathsetmacro{\r}{0.8}
\begin{tikzpicture}[baseline={([yshift=-.5ex]current bounding box.center)},every node/.style={font=\scriptsize}]
\draw [] (0,0) circle (\r cm);
\filldraw (\r,0) circle (1pt) node[right=0pt]{$5$};
\filldraw (72:\r) circle (1pt) node[above=0pt]{$4$};
\filldraw (144:\r) circle (1pt) node[above=0pt]{$3$};
\filldraw (-144:\r) circle (1pt) node[left=0pt]{$2$};
\filldraw (-72:\r) circle (1pt) node[below=0pt]{$1$};
\filldraw (\r/2,0) circle (1pt) (108:\r/2) circle (1pt) (-108:\r/2) circle (1pt);
\draw [thick] (\r,0) -- (\r/2,0) (72:\r) -- (108:\r/2) -- (144:\r) (-72:\r) -- (-108:\r/2) -- (-144:\r);
\draw [thick] (108:\r/2) -- (\r/2,0) node[pos=0.5,above right=-5pt]{};
\draw [thick] (-108:\r/2) -- (\r/2,0) node[pos=0.5,below right=-5pt]{};
\end{tikzpicture}=\frac{1}{D_{12}^{2}D_{34}^{2}}\mathcal{C}_{5} \ .
\end{split}
\end{equation}
Crucially, since $D_{I}^{2}$ and $D_{I'}^{2}$ commute if $I\subseteq I'$, $I'\subseteq I$ or $I\cap I'=\emptyset$, $D_{I}^{2}$ and $D_{I'}^{2}$ always commute if they belong to the same Witten diagram on the support of the CWI \cite{noncommuteDsubt}. Therefore, there is never any ambiguity in the ordering of the $D_{I}^{2}$ at tree level.

\section{The Differential Representation at One Loop}

We now turn to the generalization of the differential representation beyond tree level. We motivate our construction using the triangle Witten diagram 
%
\begin{align}\label{trianglpicture}
\mathcal{A}_{3}^{\triangle}=\begin{tikzpicture}[baseline={([yshift=-.5ex]current bounding box.center)},every node/.style={font=\scriptsize}]\pgfmathsetmacro{\r}{0.8}
\draw [] (0,0) circle (\r cm);
\tikzset{decoration={snake,amplitude=.4mm,segment length=1.5mm,post length=0mm,pre length=0mm}}
\filldraw (0:\r) circle (1pt) node[right=0pt]{$1$};
\filldraw (120:\r) circle (1pt) node[above=0pt]{$2$};
\filldraw (240:\r) circle (1pt) node[below=0pt]{$3$};
\filldraw (0:\r/1.7) circle (1pt) node[right=0pt]{};
\filldraw (120:\r/1.7) circle (1pt) node[above=0pt]{};
\filldraw (240:\r/1.7) circle (1pt) node[below=0pt]{};
\filldraw (60:\r/2) circle (0pt) node{};
\filldraw (180:\r/1.6) circle (0pt) node{};
\filldraw (300:\r/2) circle (0pt) node{};
\draw [thick] (0:\r) -- (0:\r/1.7);
\draw [thick] (120:\r) -- (120:\r/1.7);
\draw [thick] (240:\r) -- (240:\r/1.7);
\draw [thick] (120:\r/1.7) -- (0:\r/1.7);
\draw [thick] (240:\r/1.7) -- (120:\r/1.7);
\draw [thick] (240:\r/1.7) -- (0:\r/1.7);
\end{tikzpicture} \ .
\end{align}
Given the position space representation of $\mathcal{A}_{3}^{\triangle}$,
\begin{equation}
\begin{split}
\mathcal{A}_{3}^{\triangle}&=\int_{\textrm{AdS}}  dX_{1}dX_{2}dX_{3}  G(X_{1},X_{2})G(X_{2},X_{3}) \\
&\times G(X_{1},X_{3})\prod_{i=1}^{3}E(X_{i},P_{i}) \\
\end{split}
\end{equation}
we replace $G(X_{2},X_{3})$ with its split-representation \cite{Leonhardt:2003qu,Leonhardt:2003sn,Penedones:2007ns,Cornalba:2007fs,Cornalba:2008qf,Penedones:2010ue,Costa:2014kfa}, 
\begin{equation}\label{splitreprese}
\begin{split}
&G_{\Delta}(X_{2},X_{3})=\int_{-i\infty}^{i\infty} \frac{dc}{2\pi i}(-2c^{2})\int_{\partial \textrm{AdS}}dQdQ' \\
&\times  \frac{\delta^{d}(Q,Q')}{D_{Q}^{2}-\Delta(d-\Delta)}E_{\frac{d}{2}+c}(X_{2},Q)E_{\frac{d}{2}-c}(X_{3},Q') \ .
\end{split}
\end{equation}
with $\Delta=d$. Upon making this replacement, the triangle Witten diagram simplifies to the form 
\begin{equation}\label{preamptri1}
\begin{split}
\mathcal{A}_{3}^{\triangle}&= \int_{-i\infty}^{i\infty} \frac{dc}{2\pi i}(-2c^{2})\int_{\partial \textrm{AdS}}dQdQ'\delta(Q,Q') \\
&\times \frac{1}{D_{Q}^{2}} \mathcal{A}_{5}(P_{1},P_{2},P_{3},Q,Q')
\end{split}
\end{equation}
where $\mathcal{A}_{5}$ is a 5-point tree-level Witten diagram. So far, we have simply rewritten the loop diagram as a spectral integral over a tree diagram, as is standard \cite{Liu:2018jhs}. We now write the tree diagram in the differential representation
\begin{equation}\label{preamptri2}
\begin{split}
\mathcal{A}_{5}&=\pgfmathsetmacro{\r}{0.8}
\begin{tikzpicture}[baseline={([yshift=-.5ex]current bounding box.center)},every node/.style={font=\scriptsize}]
\draw [] (0,0) circle (\r cm);
\filldraw (\r,0) circle (1pt) node[right=0pt]{$1$};
\filldraw (72:\r) circle (1pt) node[above=0pt]{$2$};
\filldraw (144:\r) circle (1pt) node[left=0pt]{$Q'$};
\filldraw (-144:\r) circle (1pt) node[left=0pt]{$Q$};
\filldraw (-72:\r) circle (1pt) node[below=0pt]{$3$};
\filldraw (\r/2,0) circle (1pt) (108:\r/2) circle (1pt) (-108:\r/2) circle (1pt);
\draw [thick] (\r,0) -- (\r/2,0) (72:\r) -- (108:\r/2) -- (144:\r) (-72:\r) -- (-108:\r/2) -- (-144:\r);
\draw [thick] (108:\r/2) -- (\r/2,0) node[pos=0.5,above right=-5pt]{};
\draw [thick] (-108:\r/2) -- (\r/2,0) node[pos=0.5,below right=-5pt]{};
\end{tikzpicture} \\
&=\frac{1}{D_{Q3}^{2}}\frac{1}{D_{Q31}^{2}}\mathcal{C}_{5}^{c}(P_{1},P_{2},P_{3},Q,Q') \ .
\end{split}
\end{equation}
where the $c$ superscript indicates that the conformal dimensions associated with $Q$ and $Q'$ external states in Eq. (\ref{preamptri2}) are $\Delta_{Q}=d/2+c$ and $\Delta_{Q'}=d/2-c$ respectively. Combining Eqs. (\ref{preamptri1}) and (\ref{preamptri2}), we find 
\begin{equation}\label{finaltriang}
\begin{split}
\mathcal{A}_{3}^{\triangle}&= \int_{-i\infty}^{i\infty} \frac{dc}{2\pi i}(-2c^{2})\int_{\partial \textrm{AdS}}dQdQ'\delta(Q,Q') \\
&\times \frac{1}{D_{Q}^{2}}\frac{1}{D_{Q3}^{2}}\frac{1}{D_{Q31}^{2}}\mathcal{C}_{5}^{c}(P_{1},P_{2},P_{3},Q,Q') \ .
\end{split}
\end{equation}
This is the differential representation of the triangle one-loop Witten diagram. 

The above manipulations can be performed on any one-loop Witten diagram. One simply uses the split representation, Eq. (\ref{splitreprese}), to convert the one-loop, $n$-point Witten diagram to a tree-level $(n+2)$-point Witten diagram in the differential representation \cite{Qcutnote}. For example, repeating the above manipulations for bubble and box Witten diagrams, one finds
\begin{equation}
\begin{split}
\mathcal{A}^{\textrm{Bubble}}_{3}&= \int_{-i\infty}^{i\infty} \frac{dc}{2\pi i}(-2c^{2})\int_{\partial \textrm{AdS}}dQdQ'\delta(Q,Q') \\
&\times \frac{1}{D_{Q}^{2}}\frac{1}{D_{Q3}^{2}}\mathcal{C}_{5}^{c}(P_{i},Q,Q')
\end{split}
\end{equation}
and 
\begin{equation}\label{boxdiagr}
\begin{split}
\mathcal{A}^{\textrm{Box}}_{4}&= \int_{-i\infty}^{i\infty} \frac{dc}{2\pi i}(-2c^{2})\int_{\partial \textrm{AdS}}dQdQ'\delta(Q,Q') \\
&\times \frac{1}{D_{Q}^{2}D_{Q1}^{2}D_{Q12}^{2}D_{Q123}^{2}} \mathcal{C}_{6}^{c}(P_{i},Q,Q') \ .
\end{split}
\end{equation}
Notably, the first lines of Eqs. (\ref{finaltriang})-(\ref{boxdiagr}) are universal. In contrast, the second lines are unique to the Witten diagram and analogous to the corresponding Feynman diagram under the replacement of the internal loop momentum with $D_{Q}^{AB}$. 

We interpret the universal integrals over $c$, $Q$ and $Q'$ in the first lines of Eqs. (\ref{finaltriang})-(\ref{boxdiagr}) as the AdS analog of $\int dl^{\mu}$. We refer to such scalar integrals collectively as an operator-valued integral and formally define the operator-valued integral of an operator-valued integrand, $\hat{\mathcal{I}}(D_{Q},D_{i})$, as
\begin{empheq}[box=\fbox]{equation}
\begin{split}
&\int [\mathcal{D}D_{Q}]\hat{\mathcal{I}}(D_{Q},D_{i})\equiv \int_{\partial \ads} dQdQ'  \delta^{d}(Q,Q')\\
&\times  \int_{-i\infty}^{i\infty} \frac{dc}{2\pi i}(-2c^{2}) \hat{\mathcal{I}}(D_{Q},D_{i})\mathcal{C}_{n+2}^{c}(P_{i},Q,Q') \label{initialformula}  
\end{split}
\end{empheq}
where $\mathcal{C}_{n+2}^{c}(Q,Q',P_{i})$ is an $(n+2)$-point contact diagram,
\begin{equation}
\begin{split}
\mathcal{C}_{n+2}^{c}(Q,Q',P_{i})&=\int_{\ads}dX E_{\frac{d}{2}+c}(Q,X)\\
&\times E_{\frac{d}{2}-c}(Q',X)\prod_{i=1}^{n} E_{\Delta_{i}}(P_{i},X)  \ .
\end{split}
\end{equation}
Again, the $c$ superscript refers to how the conformal dimensions of the $Q$ and $Q'$ states depend on $c$. Our notation is meant to suggest that we should interpret Eq. (\ref{initialformula}) as an integral over $D_{Q}$. Using this notation, the triangle Witten diagram is
\begin{equation}\label{trianglefinal}
\begin{split}
\mathcal{A}_{3}^{\triangle}
&= \int [\mathcal{D}D_{Q}]\frac{1}{D_{Q}^{2}}\frac{1}{D_{Q3}^{2}}\frac{1}{D_{Q31}^{2}} \ ,
\end{split}
\end{equation}
and similiarly for the bubble and box differential representations. The operator-valued integrals evaluate to functions of conformal generators of external states acting on contact diagrams, $\mathcal{C}_{n}$.

The operator-valued integral notation is interesting because it simplifies expressions and provides a representation of Witten diagrams analogous to Feynman diagrams. However, the utility of the operator-valued integral goes beyond aesthetics. We show in Sections \ref{trianglexpliciteval} and \ref{genintegrationidentites} that certain identities of scalar integrals generalize to operator-valued integrals and can be leveraged to simplify the evaluation of specific Witten diagrams.

\section{Explicit Calculations at Three-Point}\label{trianglexpliciteval}

The differential representation is particularly useful for performing direct integration of one-loop Witten diagrams. This is most apparent at three-point where a number of simplifications occur, specifically a form of tensor-reduction. For Feynman integrals, tensor reduction implies that three-point, one-loop integrals obey the identity
\begin{equation}\label{feynmanampexp}
0=\int d^{d+1}l f(l^{2})(l\cdot p_{i})^{N}(l\cdot p_{j})^{M}  |_{p_{i}^{2}=0}  
\end{equation}
for any integers $N,\ M$ such that $M\geq 0$, $N\geq 0$ and $N+M>0$ \cite{tensorreductionnote}. For three-point Witten diagrams, we conjecture an analogous identity holds if $\Delta=d$ for all external states:
\begin{empheq}[box=\fbox]{equation}\label{linearchangeofvariables}
\begin{split}
&0=\int [\mathcal{D}D_{Q}]\hat{f}(D_{Q}^2)(D_{Q}\cdot D_{i})^{N}(D_{Q}\cdot D_{j})^{M}|_{n=3}
\end{split}
\end{empheq}
with the same conditions on $N$ and $M$, for all possible orderings of the differential operators in the integrand. Eq. (\ref{linearchangeofvariables}) is much more non-trivial than its flat-space analog. Even if one assumes tensor reduction is applicable to operator-valued integrals, conformal generators can in principle be contracted using the structure constants of the AdS isometry group as well as dot products. Using formulas in Appendix \ref{usefufulidentitiesss}, we explicitly checked Eq. (\ref{linearchangeofvariables}) holds for $N+M\leq 10$. In Appendix \ref{proofformua}, we prove Eq. (\ref{linearchangeofvariables}) for the special case that $N=0$. 

Eq. (\ref{linearchangeofvariables}) can be leveraged to dramatically simplify the calculation of certain three-point Witten diagrams. As an illustrative example, consider the three-point bubble diagram:
\begin{align}
\mathcal{A}_{3}^{\textrm{Bubble}}&= \begin{tikzpicture}[baseline={([yshift=-.5ex]current bounding box.center)},every node/.style={font=\scriptsize}]\pgfmathsetmacro{\r}{0.8}
\draw [] (0,0) circle (\r cm);
\draw [thick] (0,0) circle (0.3 cm);
\tikzset{decoration={snake,amplitude=.4mm,segment length=1.5mm,post length=0mm,pre length=0mm}}
\filldraw (200:\r) circle (1pt) node[left=0pt]{$1$};
\filldraw (160:\r) circle (1pt) node[left=0pt]{$2$};
\filldraw (0:\r) circle (1pt) node[right=0pt]{$3$};
\filldraw (90:0.55 cm) circle (0pt) node{$\Delta_{l}$};
\filldraw (270:0.55 cm) circle (0pt) node{$\Delta_{l}$};
\filldraw (180:0.3 cm) circle (1pt) node[above=0pt]{};
\filldraw (0:0.3 cm) circle (1pt) node[below=0pt]{};
\draw [thick] (200:\r) -- (180:0.3 cm);
\draw [thick] (160:\r) -- (180:0.3 cm);
\draw [thick] (0:\r) -- (0:0.3 cm);
\end{tikzpicture} 
\end{align}
where the conformal dimension of the state running in the loop, $\Delta_{l}$, is left unfixed. We restrict this computation to $d=2$ as this Witten diagram diverges for $d\geq 3$. The differential representation of $\mathcal{A}^{\textrm{Bubble}}_{3}$ is 
\begin{equation}\label{initialintegrandd}
\int [\mathcal{D}D_{Q}] \frac{1 }{(D_{Q}^{2}-\Delta_{l}(d-\Delta_{l}))(D_{Q3}^{2}-\Delta_{l}(d-\Delta_{l}))} \ . 
\end{equation}
Since $\Delta_{3}=d=2$, we find that $D_{3}^{2}=0$. Performing a Taylor Series in $D_{Q}\cdot D_{3}$, one finds that all terms vanish due to Eq. (\ref{linearchangeofvariables}) except the leading term. Therefore, the bubble Witten diagram simplifies to
\begin{equation}\label{seconint}
\int [\mathcal{D}D_{Q}]\frac{1}{(D_{Q}^{2}-\Delta_{l}(2-\Delta_{l}))^{2}} \ .
\end{equation}
Substituting the definition of the operator-valued integral and using \cite{Penedones:2010ue,Simmons-Duffin:2012juh}
\begin{equation}\label{goodfunctionpartial}
\begin{split}
&\frac{\Gamma(d/2)\Gamma(d/2+c)\Gamma(d/2-c)}{4\pi^{d/2}\Gamma(d)\Gamma(1-c)\Gamma(1+c)}\\
&=\int_{\partial \ads}dQ E_{d/2+c}(Q,X)E_{d/2-c}(Q,X) \ ,   
\end{split}
\end{equation}
we reduce the integral to a single contour integral which can be evaluated using the residue theorem. The final result for $\mathcal{A}_{3}^{\textrm{Bubble}}$ is 
\begin{equation}\label{finalresult}
\begin{split}
&\int_{-i\infty}^{i\infty} \frac{dc}{(2\pi)^{2} i}\frac{\Gamma(1+c)\Gamma(1-c)\mathcal{C}_{3}}{ \Gamma(c)\Gamma(-c)(1-c^{2}-\Delta_{l}(2-\Delta_{l}))^{2}} \\
&=\frac{1}{8\pi (\Delta_{l}-1)}\mathcal{C}_{3}\ .
\end{split}
\end{equation}
This result is cross-checked in Appendix \ref{directcheckbubble}, where we evaluate the bubble Witten diagram in position space and find the answers agree. 

We can use the differential representation to evaluate more complex Witten diagrams, such as the triangle Witten diagram. We fix the conformal dimension of states running in the loop to $\Delta_{l}=d$ for simplicity. The relevant operator-valued integral is then Eq. (\ref{trianglefinal}). We again take a Taylor series of the operator-valued integrand, except now in $D_{Q}\cdot D_{1}$ and $D_{Q}\cdot D_{2}$. All terms vanish except the leading term due to Eq. (\ref{linearchangeofvariables}). The final result can be converted into a single scalar integral, which can again be evaluated using residue theorem. Evaluating the integral, we found
\begin{equation}
\begin{split}
\mathcal{A}_{3}^{\triangle}|_{d=2}&=\frac{1}{32\pi
}\mathcal{C}_{3}, \quad \mathcal{A}_{3}^{\triangle}|_{d=4}=\frac{13}{1536\pi^{2}}\mathcal{C}_{3}  \ , \\
\mathcal{A}_{3}^{\triangle}|_{d=3}&=\frac{7 \pi ^2-36 \zeta (3)-6}{1296 \pi ^2}\mathcal{C}_{3} \ ,
\end{split}
\end{equation}
and that the integral is divergent for $d\geq 5$, similar to flat space. Evaluating the $c$-integral for odd $d$ is slightly harder than even $d$ because an infinite number of residues contribute that need to be re-summed.

\section{Generalized IBP Relations}\label{genintegrationidentites}

In flat space, IBP is an important tool for computing Feynman integrals \cite{Chetyrkin:1981qh,vonManteuffel:2012np,vonManteuffel:2014ixa,Maierhofer:2017gsa,Mastrolia:2018uzb,Smirnov:2019qkx,Frellesvig:2019kgj,Frellesvig:2019uqt,Mizera:2019vvs,Mizera:2019ose,Frellesvig:2020qot,Laporta:2000dsw}. We now give a partial generalization of IBP for operator-valued integrals. We first note that the operator valued integral should be invariant under arbitrary conformal transformations of $Q$ and $Q'$, which implies
\begin{equation}
\mathcal{I}=\int [\mathcal{D}D_{Q}]e^{v\cdot (D_{Q}+D_{Q'})}\hat{\mathcal{I}} \ ,
\end{equation}
where $v$ is a tensor, is independent of $v$. We now rewrite the above operator-valued integral as 
\begin{equation}\label{orgintegral}
\mathcal{I}=\int [\mathcal{D}D_{Q}]\hat{\mathcal{I}}'e^{-v\cdot (\sum_{i=1}^{n}D_{i})}
\end{equation}
where $\hat{\mathcal{I}}'$ is $\hat{\mathcal{I}}$ with the replacement
\begin{equation}\label{nonlineashift}
\begin{split}
D_{a}^{AB}\rightarrow e^{v\cdot (D_{Q}+D_{Q'})}D_{a}^{AB}e^{-v\cdot (D_{Q}+D_{Q'})} \ .
\end{split}
\end{equation}
for all $a\in \{Q,1,\ldots,n\}$. If $v$ is a constant tensor, then the above shift only acts non-trivially on $D_{Q}$ and dependence on $D_{Q'}$ disappears. Let us now take $v$ to be an infinitesimal in Eq. (\ref{orgintegral}). Since the result is independent of $v$, the component linear in $v$ must vanish, which imposes non-trivial linear relations among operator-valued integrals. The collection of identities derivable from this procedure does not necessarily span the space of all linear identities obeyed by operator-valued integrals, but is enough to illustrate that there are non-trivial relations which mimic their flat-space counter-parts.

For example, we can apply the above procedure to the triangle Witten diagram. We assume $v$ is an infinitesimal constant, so the replacement rule simplifies to 
\begin{equation}
\begin{split}
D_{Q}^{AB}&\rightarrow  D_{Q}^{AB}+f^{AB}_{CD,EF}D_{Q}^{CD}v^{EF} \ . 
\end{split}
\end{equation}
where $f^{AB}_{CD,EF}$ is a structure constant of the AdS isometry group. The above procedure ultimately implies the operator-valued integrand
\begin{equation}\label{IBPnontrivialidentit}
\begin{split}
&\hat{\mathcal{I}}=\frac{v_{AB}f^{AB}_{CD,EF}}{D_{Q}^{2}D_{Q1}^{2}D_{Q12}^{2}}\Bigg ((D_{Q}^{CD}D_{1}^{EF})\frac{1}{D_{Q1}^{2}} \\
+&(D_{Q}^{CD}D_{12}^{EF})\frac{1}{D_{Q12}^{2}}\Bigg)-\frac{1}{D_{Q}^{2}D_{Q1}^{2}D_{Q12}^{2}}(v\cdot \sum_{i=1}^{3}D_{i}),
\end{split}    
\end{equation}
integrates to zero for external states with arbitrary conformal dimension. Unlike the operator-valued integrands previously considered, the differential operators in each term do not always commute and there are contractions of conformal generators with structure constants. Furthermore, the constant tensor $v$ explicitly breaks conformal symmetry, so the CWI must be applied with care \cite{conformalbreaking}.

\section{Outlook}

The differential representation is a powerful framework for evaluating Witten diagrams, as illustrated by the direct evaluation of the triangle Witten diagram. Beyond three-point, the differential representation implies linear relations among certain operator-valued integrands, which are the AdS generalization of IBP relations. In general, the similarities between Witten diagrams in the differential representation and Feynman diagrams imply that many techniques for evaluating Feynman diagrams should generalize to the differential representation. Beyond AdS, the differential representation can also be used to evaluate Witten diagrams in de Sitter space \cite{Maldacena:2002vr,Arkani-Hamed:2015bza,Arkani-Hamed:2018kmz,Gomez:2021qfd,Baumann:2021fxj,Hillman:2021bnk,dScommentradiIR}.

\begin{acknowledgements}


AH is grateful to Clifford Cheung, Henriette Elvang, Sebastian Mizera, Julio Parra Martinez, Radu Roiban, Allic Sivaramakrishnan, and Fei Teng for stimulating discussion and insight. AH is supported in part by the US Department of Energy under Grant No. DE-SC0007859 and in part by a Leinweber Center for Theoretical Physics Graduate Fellowship. 
\end{acknowledgements}

\appendix

\section{Conventions}\label{concventionsapp}

We now review the convention choices of this letter. Although the differential representation is equally applicable in momentum and position space \cite{Gomez:2021qfd}, we work in position space, specifically the embedding representation, when comparing the differential representation to more familiar expressions. Since we do not study the Witten diagrams of higher spin states in the main text, only conventions relevant for evaluating scalar Witten diagrams are discussed.

Under Cartesian coordinates $X^A=(X^{\mathsf{a}},X^d,X^{d+1})$ and setting the $\ads$ lenth scale to one, the $\ads_{d+1}$ hypersurface corresponds to fixing $X^{2}=-1$. The Poincare coordinates of the $\ads_{d+1}$ hypersurface in $\mathbb{R}^{d+1,1}$ are given by 
\begin{equation}
    \begin{split}
\label{eq:embedding}
     X^{\mathsf{a}}&=\frac{1}{z}x^{\mathsf{a}}\,, \\
     X^d&=\frac{1}{z}\frac{1-x^2-z^2}{2}\,, \\
     X^{d+1}&=\frac{1}{z}\frac{1+x^2+z^2}{2}\,,
    \end{split}
\end{equation}
such that
\begin{align}\label{eq:metricAdS}
    ds^2_{\ads_{d+1}}=\frac{1}{z^2}\left(dz^2+dx_{\mathsf{a}} dx^{\mathsf{a}}\right)\,.
\end{align}
The boundary is given by the projective null cone in $\mathbb{R}^{d+1,1}$, 
\begin{align}
    P_i = \Big(x_i^{\mathsf{a}},\frac{1-x_i^2}{2},\frac{1+x_i^2}{2}\Big)\,,
\end{align}
such that $P^{2}=0$ and $P_{i}^{A}\sim \lambda P_{i}^{A}$. Conformal generators in the embedding space representation are given in Eq. (\ref{conformalgenerat}) in the main text. In our chosen normalization, the commutator of two conformal generators is
\begin{equation}
\begin{split}
&[D_{i}^{AB},D_{j}^{CD}]=\delta_{ij}f^{AB,CD,EF}D_{i,EF} \\
\end{split}
\end{equation}
where $f^{AB,CD,EF}$ is the $SO(d,2)$ structure constant
\begin{equation}
\begin{split}
f^{AB,CD,EF}&=\frac{1}{\sqrt{2}}[\eta^{BC}\eta^{AE}\eta^{DE}-(A\leftrightarrow B)] \\
&-(C\leftrightarrow D) \ .
\end{split}
\end{equation}
Finally, we need to define a metric tensor that projects dynamics onto the embedding hypersurface:
\begin{equation}
G_{AB}=\eta_{AB}-\frac{X_{A}X_{B}}{X^{2}}=g^{\mu\nu}\frac{\partial X_{A}}{\partial x^{\mu}}\frac{\partial X_{B}}{\partial x^{\nu}} .    
\end{equation}
Crucially, $G^{AB}$ is defined such that 
\begin{equation}
\nabla_{\ads}^{2}=\partial_{A}\left (G^{AB}\partial_{B}\right )
\end{equation}

We now turn to propagators. We choose to normalize the bulk-to-bulk propagators such that 
\begin{equation}\label{normalizationbulkbulk}
\big[\, -D_{X_{1}}^{2}- \Delta(\Delta-d)\big]G_{\Delta}(X_{1},X_{2})=-\delta^{d+1}(X_{1},X_{2})
\end{equation}
where one can show $D_{X}^{2}=\partial_A (G^{AB}\partial_B)$. In terms of hypergeometric functions, the bulk-to-bulk propagator, $G_{\Delta}(X_{1},X_{2})$, is
\begin{equation}
\begin{split}
\frac{C_{\Delta}}{(u)^{\Delta}}{}_{2}F_{1}\left (\Delta,\frac{2\Delta-d+1}{2},2\Delta-d+1,\frac{-4}{u} \right ) \ ,    
\end{split}
\end{equation}
where $C_{\Delta}$ is a normalization factor, 
\begin{equation}
C_{\Delta}=\frac{\Gamma(\Delta)}{2\pi^{d/2}\Gamma(\Delta-d/2+1)}    
\end{equation}
and $u$ is the chordal distance,
\begin{equation}
u=(X_{1}-X_{2})^{2}\ .
\end{equation}
The bulk-to-boundary propagator is defined in terms of the asymptotic limit of the bulk-to-bulk propagator,
\begin{equation}\label{normbulkbound}
\begin{split}
E_{\Delta}(X_{1},X_{2})&=\lim_{X_{1}\rightarrow \partial\ads}z_{1}^{-\Delta}G_{\Delta}(X_{1},X_{2})   \\
&=\frac{C_{\Delta}}{(-2X_{1}\cdot X_{2})^{\Delta}} \ .
\end{split}
\end{equation}
Again, the normalization conventions of the propagators are fully fixed by Eqs. (\ref{normalizationbulkbulk}) and (\ref{normbulkbound}).

Finally, AdS propagators can be written as superpositions of AdS harmonic functions. AdS harmonic functions are defined by the differential equation
\begin{equation}
(D_{X_{1}}^{2}-(d^{2}/4-c^{2}))\Omega_{c}(X_{1},X_{2})=0  \ ,
\end{equation}
and normalized such that 
\begin{equation}
\delta^{d+1}(X_{1},X_{2})=\int_{-i\infty}^{i\infty} \frac{dc}{2\pi i}  \Omega_{c}(X_{1},X_{2}) \ .
\end{equation}
Crucially, AdS harmonic functions satisfy an orthogonality condition 
\begin{equation}\label{orthogonalityconditoi}
\begin{split}
&\int_{\ads} dX_{2}\Omega_{c}(X_{1},X_{2})\Omega_{c'}(X_{2},X_{3})\\
&=\pi [\delta(c-c')+\delta(c+c')]\Omega_{c}(X_{1},X_{3}) \ ,
\end{split}
\end{equation}
which is very useful when integrating sequences of propagators. In terms of AdS harmonic functions, the AdS bulk-to-bulk propagator is 
\begin{equation}\label{bulktobulkrepresention}
G_{\Delta}(X_{1},X_{2})=\int_{-i\infty}^{i\infty} \frac{dc}{2\pi i}\frac{\Omega_{c}(X_{1},X_{2})}{(\Delta-d/2)^{2}-c^{2}}  \ .  
\end{equation}
We can invert Eq. (\ref{bulktobulkrepresention}) to find a representation of $\Omega_{c}(X_{1},X_{2})$ in terms of bulk-to-bulk propagators,
\begin{equation}\label{linearrep}
\begin{split}
&\Omega_{c}(X_{1},X_{2}) \\
&\quad =c(G_{d/2+c}(X_{1},X_{2})-G_{d/2-c}(X_{1},X_{2}))  \ .  
\end{split}
\end{equation}
A particularly useful representation of $G_{\Delta}(X_{1},X_{2})$ comes from substituting the split representation of the harmonic function,
\begin{equation}\label{splitrep}
\begin{split}
&\Omega_{c}(X_{1},X_{2}) \\
&\quad =-2c^{2}\int dQ E_{d/2+c}(Q,X_{1})E_{d/2-c}(Q,X_{2})  \ ,
\end{split}
\end{equation}
into Eq. (\ref{bulktobulkrepresention}):
\begin{equation}\label{splitrepresegen}
\begin{split}
&G_{\Delta}(X_{2},X_{3})=\int_{-i\infty}^{i\infty} \frac{dc}{2\pi i}(-2c^{2})\int_{\partial \textrm{AdS}}dQdQ' \\
&\times \frac{\delta^{d}(Q,Q') }{D_{Q}^{2}-\Delta(d-\Delta)}E_{\frac{d}{2}+c}(X_{2},Q)E_{\frac{d}{2}-c}(X_{3},Q') \ .
\end{split}
\end{equation}
Further identities are provided in the main text as needed. 

\section{Useful Integral Identities}\label{usefufulidentitiesss}

In this appendix, we discuss how we checked that Eq. (\ref{linearchangeofvariables}) holds. Crucially, first note that $\hat{f}(D_{Q}^{2})$ becomes $\hat{f}(d^{2}/4-c^{2})$ upon acting on $\mathcal{C}_{5}(P_{i},Q,Q')$ in Eq. (\ref{initialformula}) and is therefore independent of $Q$. However, non-trivial dependence on $Q$ emerges upon acting $(D_{Q}\cdot D_{i})^{N}(D_{Q}\cdot D_{j})^{M}$ on $\mathcal{C}_{5}(P_{i},Q,Q')$. We explicitly checked that the resulting expression vanishes upon integrating over $Q$ for $N+M\leq 10$. 

We review the computation strategy to integrate over $Q$. We find that the integrand contains terms whose $Q$-dependence takes the generic form
\begin{equation}\label{orgformaul}
    I_{a_{1},a_{2},\ldots}=\int_{\partial \ads} dQ \frac{\prod_{i}(-2Q\cdot P_{i})^{a_{i}}}{(-2Q\cdot X)^{d+\sum_{i}a_{i}}} \ .    
\end{equation}
We first consider the simplest specialization of Eq. (\ref{orgformaul}):
\begin{equation}\label{preintegratiuon}
    I_{a}=\int_{\partial \ads} dQ \frac{(-2Q\cdot P_{1})^{a}}{(-2Q\cdot X)^{d+a}} \ .
\end{equation}
Using the identity 
\begin{equation}
\frac{\Gamma[a]}{f^{a}}=\int^{\infty}_{0}\frac{dv}{v}v^{a}e^{-vf} \ ,
\end{equation}
we can rewrite Eq. (\ref{preintegratiuon}) as 
\begin{equation}\label{preintegratiuon2}
\left ( \frac{\partial}{\partial \alpha} \right )^{a}\int_{\partial \ads} dQ\int^{\infty}_{0}\frac{dv}{v}\frac{v^{d+a}e^{2Q\cdot(vX-\alpha P_{1})}}{\Gamma[d+a]}|_{\alpha=0} \ .
\end{equation}
Finally, using the identity
\begin{equation}
\int_{\partial \ads}dQ e^{2Q\cdot T}=\frac{\pi^{d/2}}{|T|^{d/2}}e^{-|T|}
\end{equation}
the integral over $Q$ in Eq. (\ref{preintegratiuon2}) yields
\begin{equation}
\left ( \frac{\partial}{\partial \alpha} \right )^{a}\int \frac{ v^{d+a-1}\pi^{d/2}e^{-\sqrt{v\alpha(-2X\cdot P_{1})-v^{2}}}   dv}{\Gamma[d+a](v\alpha(-2X\cdot P_{1})-v^{2})^{d/4}}|_{\alpha=0}
\end{equation}
which simplifies to 
\begin{equation}\label{easyresult}
I_{a}=(-2 P_{1}\cdot X)^{a}\pi^{d/2}\frac{\Gamma[\frac{d}{2}+a]}{\Gamma[d+a]} \ .
\end{equation}
This computation strategy generalizes to all integrals of the form Eq. (\ref{orgformaul}). Writing the result in a tensor version of Eq. (\ref{orgformaul}), the integral yields
\begin{equation}\label{finalformula}
\begin{split}
I^{A_{1},\ldots,A_{n}}&=\int dQ \frac{Q^{A_{1}}\ldots Q^{A_{n}}}{(-2Q\cdot X)^{d+n}} \ , \\
&=\frac{\pi^{d/2}\Gamma(d/2+n)}{\Gamma(d+n)}X^{A_{1}}\ldots X^{A_{n}}-\textrm{Traces}
\end{split}    
\end{equation}
where traces are subtracted using $\eta^{AB}$. 

Eq. (\ref{finalformula}) was originally given in Ref. \cite{Simmons-Duffin:2012juh} by taking derivatives of the integral 
\begin{equation}
\begin{split}
I(X)&=\int dQ \frac{1}{(-2Q\cdot X)^{d}} \\
&=\frac{\pi^{d/2}\Gamma(d/2)}{\Gamma(d)}\frac{1}{(-X^{2})^{d/2}} \ .
\end{split}
\end{equation}
in the bulk coordinate $X^{A}$. We have reproduced this formula here by direct integration to avoid subtleties that are relevant when taking derivatives in bulk or boundary coordinates in embedding space \cite{exampsubtder}.  


\section{Proof of Eq. (\ref{linearchangeofvariables}) when $N=0$}\label{proofformua}

In this appendix, we prove that Eq. (\ref{linearchangeofvariables}) holds for general $n$-point integrands, not just at three-point, when $N=0$ \cite{privatecorrespondenceFei}. We then sketch a possible proof strategy of Eq. (\ref{linearchangeofvariables}) for general $N$ and $M$. 

We first note that when $(D_{Q}\cdot D_{1})^{N}$ acts on $D_{\Delta_{Q},\Delta_{1},\ldots}$, the result takes the form 
\begin{equation}\label{expressiontosolve}
\begin{split}
&(D_{Q}\cdot D_{1})^{N}D_{\Delta_{Q},d\ldots} \\
&\ =\sum a_{n,k}(Q\cdot P_{1})^{k}D_{\Delta_{Q}+k,d+k,\ldots}    
\end{split}
\end{equation}
where $D_{\Delta_{Q}+k,d+k,\ldots}$ is the D-function, defined as
\begin{equation}
\begin{split}
&D_{\Delta_{Q},\Delta_{1},\ldots }= \\
&\ \int_{\ads}dX (-2X\cdot P_{1})^{-\Delta_{Q}}(-2X\cdot P_{n})^{-\Delta_{1}}\ldots   \ .
\end{split}
\end{equation}
To solve for $a_{n,k}$, we use the relation
\begin{equation}\label{recursioncoef}
\begin{split}
&(D_{Q}\cdot D_{1})[(Q\cdot P_{1})^{k}D_{\Delta_{Q}+k,d+k,\ldots}]= \\
&-8(\Delta_{Q}+k)(d+k)(Q\cdot P_{1})^{k+1}D_{\Delta_{Q}+k+1,d+k+1,\ldots}\\
&-4(\Delta_{Q}+k)(d/2+ k)(Q\cdot P_{1})^{k}D_{\Delta_{Q}+k,d+k,\ldots}
\end{split}
\end{equation}
which provides a recursion relation for the $a_{n,k}$ coefficients,
\begin{equation}\label{aknrec}
a_{n,k}=a_{n-1,k-1}f_{k-1}+a_{n-1,k}g_{k}    
\end{equation}
where 
\begin{equation}\label{fulldefinitionfg}
    \begin{split}
        f_{k}&=-8 (\Delta_{Q}+ k) (d + k) \ ,\\
        g_{k}&=-4(\Delta_{Q} + k)(d/2 + k)  \ , \\
        a_{0,0}&=1 \ , \\
        a_{n,k}&=0 \textrm{ if }k>n\textrm{ or }k<0 \ .
    \end{split}
\end{equation}
We now take the expression in Eq. (\ref{expressiontosolve}) and integrate over $Q$ using Eq. (\ref{easyresult}). We find the result
\begin{equation}\label{expressiontosolve}
\begin{split}
&\int_{\ads} dQ \sum_{k=0}^{n} a_{n,k}(Q\cdot P_{i})^{k}D_{\Delta_{Q}+k,d+k,\ldots}   \\
&=\pi^{d/2}D_{\Delta_{Q},d,\ldots}\sum_{k=0}^{n} \left ( \frac{1}{2} \right )^{-k}a_{n,k} \frac{\Gamma(d/2+k)}{\Gamma(d+k)} \ .
\end{split}
\end{equation}    
To show this expression is zero, substitute the identity in Eq. (\ref{aknrec}) for all $a_{n,k}$. There are now two sums over $g_{k}\times (\ldots )$ and $f_{k-1}\times (\ldots )$ respectively. Substituting in the definitions of $g_{k}$ and $f_{k}$ in Eq. (\ref{fulldefinitionfg}), these two sums cancel. Therefore, the expression in Eq. (\ref{expressiontosolve}) vanishes.

Unfortunately, proving Eq. (\ref{linearchangeofvariables}) for non-zero $N$ and $M$ is much more difficult than the $N=0$ case. We will sketch a proof strategy here. Similar to the $N=0$ case, one would first establish an ansatz for $(D_{Q}\cdot D_{1})^{N}(D_{Q}\cdot D_{2})^{M}D_{\Delta_{Q},d,d,\ldots}$ as a sum of terms of the form 
\begin{equation}
\begin{split}
&(Q\cdot P_{1})^{k_{1}}(Q\cdot P_{2})^{k_{2}}(P_{1}\cdot P_{2})^{k_{3}} \\
&\times D_{\Delta_{Q}+k_{1}+k_{2},d+k_{1}+k_{3},d+k_{2}+k_{3},\ldots} \ .    
\end{split}
\end{equation}
One would then establish a recursion relation among coefficients similar to Eq. (\ref{recursioncoef}) and perform an integral over $Q$ using Eq. (\ref{finalformula}). Unlike the $N=0$ case, one would also need to subsequently integrate over the bulk coordinate $X$ using the closed form expression of the 3-point D-function. After integrating over $Q$ and $X$, the hope is that the recursion relations between coefficients would be enough to show that the terms in the sum cancel among themselves, similar to what happens in the $N=0$ case.

\section{Explicit Comparison for Bubble Diagram in $\ads_{3}$}\label{directcheckbubble}

In this appendix, we evaluate the three-point bubble diagram in Eq. (\ref{initialintegrandd}) in position space as a cross-check of our result in Section \ref{trianglexpliciteval}. To simplify the computation, we consider the more general case that $P_{3}^{A}$ is in the bulk and then take the limit that $P_{3}^{A}$ approaches the boundary, writing 
\begin{equation}\label{initialintegrand}
\begin{split}
\mathcal{A}_{3}^{\textrm{Bubble}}&=\lim_{P_{3}\rightarrow \partial\ads}(z_{3})^{-2}\int_{\ads} dX_{1}dX_{2}E(P_{1},X_{1}) \\
&\times E(P_{2},X_{1})(G_{\Delta_{l}}(X_{1},X_{2}))^{2}G(X_{2},P_{3})  \ .  
\end{split}
\end{equation}
We consider the split-representation of the $d=2$ bulk-to-bulk propagator, given in Eq. (\ref{bulktobulkrepresention}), and the bubble,
\begin{equation}\label{prebubble}
\begin{split}
G_{\Delta_{l}}(X_{1},X_{2})^{2}=\int^{i\infty}_{-i\infty}\frac{dc}{2\pi i}B_{c}^{\Delta_{l}}\Omega_{c}(X_{1},X_{2})   \ , 
\end{split}
\end{equation}
where $B(c)$ was derived in Ref. \cite{Carmi:2018qzm},
\begin{equation}\label{explicitbuuble}
B_{c}^{\Delta_{l}}=\frac{\psi(\Delta_{l}-\frac{1+c}{2})-\psi(\Delta_{l}-\frac{1-c}{2})}{8\pi c}    \ .
\end{equation}
Using orthogonality of AdS conformal partial waves, we find that 
\begin{equation}\label{finalsplitrepresention}
\begin{split}
&\int_{\ads} dX_{2} (G_{\Delta_{l}}(X_{1},X_{2}))^{2}G(X_{2},P_{3}) \\
&\quad =\int^{i\infty}_{-i\infty}\frac{dc}{2\pi i} \frac{B_{c}^{\Delta_{l}}}{1-c^{2}}\Omega_{c}(X_{1},P_{3})  \ .    
\end{split}
\end{equation}
Substituting Eq. (\ref{finalsplitrepresention}) into Eq. (\ref{initialintegrand}) and rewriting the conformal partial wave as a sum of $G_{d/2\pm c}(X_{1},X_{2})$ as in Eq. (\ref{linearrep}), the one-loop correlator simplifies to
\begin{equation}\label{finalresultbubble}
\begin{split}
&\lim_{z_{-3}\rightarrow 0}z_{3}^{-2} \int^{i\infty}_{-i\infty}\frac{dc}{2\pi i} \frac{c B_{c}^{\Delta_{l}}}{1-c^{2}}\int dX_{1}E(P_{1},X_{1})\\
& E(P_{2},X_{1})(G_{1+c}(X_{1},P_{3})-G_{1-c}(X_{1},P_{3}))
\end{split}
\end{equation}
This integral can be evaluated using the residue theorem, but the contour is different for each term due to distinct behavior at $|z|\rightarrow \infty$. The $G_{1\pm c}(X_{1},P_{3})$ term corresponds to a contour which includes the residue at $c=\pm 1$. The final result is
\begin{align}\label{finalresultdirectint}
\mathcal{A}^{\textrm{Bubble}}_{3}=\frac{1}{8\pi (\Delta_{l}-1)}\mathcal{C}_{3} \, . 
\end{align}
As expected, we find that the operator-valued integration result in Eq. (\ref{finalresult}) matches the result derived from direct integration in position space in Eq. (\ref{finalresultdirectint}). Given that $\mathcal{A}_{3}^{\textrm{Bubble}}$ is a one-loop diagram in $\ads_{3}$, it was surprisingly straightforward to evaluate.  The key to the above computation was using the split representation of the bubble diagram in Eqs. (\ref{prebubble}) and (\ref{explicitbuuble}). Unfortunately, this computation strategy does not generalize to more complicated one-loop Witten diagrams, such as $\mathcal{A}_{3}^{\triangle}$.


\bibliography{apssamp}

\begin{thebibliography}{95}
\bibitem [{non()}]{noncommuteDsubt}%
  \BibitemOpen
  \href@noop {} {}\bibinfo {note} {You can technically choose your $D_{I}^{2}$
  such that they do not commute in the same Feynman diagram. However, the
  commutator always vanishes on the support of the CWI. For example, we could
  write Eq. (\ref{5point}) using $D_{345}^{2}$ and $D_{123}^{2}$, which do not
  commute. However, the commutator of $D_{345}^{2}$ and $D_{123}^{2}$ is
  proportional to $\sum_{i=1}^{5}D_{i}$ and therefore vanishes due to the
  CWI.}\BibitemShut {Stop}%
\bibitem [{ten()}]{tensorreductionnote}%
  \BibitemOpen
  \href@noop {} {}\bibinfo {note} {Tensor reduction implies that for the
  integral in Eq. (\ref{feynmanampexp}) to not vanish, it must be proportional
  to $p_{i}\cdot p_{j}$. However, $p_{i}\cdot p_{j}$ vanishes for massless
  external kinematics at three-point.}\BibitemShut {Stop}%
\bibitem [{exa()}]{exampsubtder}%
  \BibitemOpen
  \href@noop {} {}\bibinfo {note} {For example, although the function
  $P_{1}^{2}f(P_{i})$ is zero everywhere as boundary correlators are only
  defined when $P_{i}^{2}=0$, its derivative is non-zero. Ignoring this
  subtlety can lead to contradictions when using derivatives to find identities
  among D-functions.}\BibitemShut {Stop}%
\end{thebibliography}
	
\end{document}